\documentstyle[12pt,preprint,prl,aps]{revtex}

\begin{document}

\draft

\title{Can Galactic Observations Be Explained by a
Relativistic Gravity Theory?}

\author{Vadim V. Zhytnikov\cite{ZWaddr} and James M. Nester}

\address{Department of Physics,
National Central University, Chung-Li, Taiwan 32054}

\date{June 1994}
\maketitle

\begin{abstract}
We consider the possibility of an alternative
gravity theory explaining the dynamics of galactic systems without
dark matter.  From very general assumptions about
the structure of a relativistic gravity theory we derive a general
expression for the metric to order $(v/c)^2$.  This allows
us to compare the predictions of the theory with various experimental
data: the Newtonian limit, light deflection and retardation,
rotation of galaxies and gravitational lensing.  Our general
conclusion is that the possibility for any gravity theory to explain
the behaviour of galaxies without dark matter is rather improbable.
\end{abstract}


\narrowtext

Einstein's gravity theory and some other alternative gravity
models are in good agreement with the experimental data
in the solar system and the laboratory \cite{1}.
However, the behavior of galactic systems poses a great challenge
to gravity theories.
For virtually all spiral galaxies the tangential rotational velocity
curves tend toward some constant value. This fact is in sharp
contradiction with the visible star (luminosity) distribution
and the laws of Newtonian dynamics.  The stars in the outer parts of
galaxies rotate several times faster than predicted by
the standard gravity theory.
A similar problem is observed in gravitational lensing \cite{2}.
Just as in the solar system problems of the past century
(concerning the orbits of Uranus and Mercury),
there are two ways to resolve these difficulties.

One, the most widely adopted,
is the {\it dark matter} hypothesis \cite{3}.
It is presumed that
the visible stars are imbedded in a massive nearly spherical
halo of nonluminous matter.
The mass of the halo varies from one
galaxy to another but generally it constitutes about 90\%
of the total mass.  This hypothesis explains
the flat rotational curves
of galaxies.  Yet it has its own troubles,
in particular (i) no good
model for the formation of the dark
halo is known, and (ii) after much
effort and many proposals no known form of matter has
yet given a
satisfactory model for the massive halo.
(The few recently observed cases of gravitational
microlensing \cite{4} are as yet far from conclusive evidence
for the dark matter explanation).

The second way is to assume that for galactic distances
Newton's gravity law is no longer valid.
This possibility has also been the subject of some
discussions \cite{5,6,7,8}.  In particular
it was shown \cite{9} that a modified gravitational potential
of the form
\begin{equation}
\varphi=\frac{-GM}{r(1+\alpha)}
\left[1+\alpha e^{-r/r_0}\right],
\label{sanders} \end{equation}
where $\alpha=-0.9,r_0\approx30{\rm kpc}$
can explain flat rotational curves for most of the galaxies.
The potential (\ref{sanders}) differs from the usual one by
an extra exponential term.
For the solar system this term equals 1 with high accuracy
and (\ref{sanders}) reduces to the standard form $\varphi=-GM/r$
but for distances significantly greater than $30{\rm kpc}$
the exponential term vanishes and we have once again a Newtonian
potential $\varphi=-GM/(1+\alpha)r$ but now with an approximately
ten times bigger gravitational constant $G/(1+\alpha)$.
Besides (\ref{sanders}) several other modified gravitational
models were considered in the literature \cite{5}.
Most were introduced purely phenomenologically
without derivation from some gravity theory.
All of these models and also various attempts to construct
nonrelativistic gravity theories \cite{5} have
the same trouble.  They cannot describe the motion of light
without additional assumptions.  A description of light
requires a {\em relativistic} gravity model.

We investigate the
general possibility of constructing a
relativistic gravity theory which can
explain galactic mysteries and other
experimental data like the classical
solar system tests \cite{1}.
We formulate some very general
postulates about the structure
of the theory:

(i)  Gravitational  phenomena are described by the metric of
space-time $g_{\mu\nu}$ and possibly some other set of fields
$\Psi_A$.  The theory is invariant under general coordinate
transformations.

(ii) The trajectories of (structureless) massive test particles and
light are timelike and null geodesics of the metric
$g_{\mu\nu}$ respectively.

(iii) The source for the gravitational fields
$g_{\mu\nu}$ and $\Psi_A$ are the energy-momentum tensor
$T_{\mu\nu}$ and some current $J_A$.
For the solar system and galaxies these sources can be
taken in the form $T_{\mu\nu}=T_{00}=\rho,$ $J_A=0$
in the $(v/c)^2$ approximation.

(iv) The theory has a good linear approximation.

(v) Flat spacetime
$g_{\mu\nu}=\eta_{\mu\nu}={\rm diag}(-1,1,1,1)$ and
$\Psi_A=\Psi_A^0$ ($\Psi^0_A$ some constant or
almost constant field) can be considered as the background
field configuration for the solar system and galactic scales.

(vi) The theory does not possess any unusual gauge
freedom for the metric field besides general coordinate
invariance.  Any gauge freedom for the field $\Psi_A$ is
fixed by an appropriate gauge fixing condition.

(vii) The theory is not a higher-derivative theory.

Let us briefly discuss postulates (i)--(vii).
The postulates (i),(ii) are the usual postulates
of the so called {\em metric theory of gravity} \cite{1}.
Postulate (iii), especially the condition
$J_A=0$, is of crucial importance for our study.
First, it allows us to make {\em definite} predictions
about the post-Newtonian approximation of the theory
without needing detailed information about its structure.
Second, as we'll see below, it ensures that
the theory does not violate the equivalence principle.
The nature of the current $J_A$ may be different;
for some models it may be
absent explicitly. In particular,
for the Brans--Dicke \cite{1}
theory $\Psi_A$ is the scalar field and it
does not have any corresponding matter source.  However, for the
Poincar\'e gauge theory of gravitation \cite{10}
$\Psi_A$ is the space-time torsion and
the current is the spin-tensor of matter which
vanishes to a high approximation since both
the solar system and the galaxies do not contain
large amounts of spin-polarized matter.
Postulate (iv) excludes from our considerations
all essentially nonlinear theories for which
the linear approximation is invalid.
Postulate (v) means that we neglect
global cosmological effects.
Postulate (vi) ensures that gravitational
equations are nondegenerate. The assumption about
the absence of extra gauge freedom for the metric is quite
natural since such invariance normally imposes
unphysical constraints on the energy-momentum
tensor of matter fields.
Postulate (vii) excludes from our consideration theories
with Green functions of
the form $1/(\Box-m^2)^n,n\geq2$.
These postulates are actually not too restrictive.
For example they
allow a large class of geometric gravity
theories derived from Lagrangians
depending on the metric and the connection
through the torsion,
curvature and non-metricity.

We are going to compare the predictions of any gravity theory
which satisfies postulates (i)--(vii) with the
experimental data from the solar system and the galaxies.
A key point is that all of these systems are essentially post-Newtonian
slow-motion, weak-gravitational-field systems \cite{1}.
Hence we can consider our theory in the linearized approximation
and we can use the small post-Newtonian parameter
$\varphi\approx v^2\ll1$ (we assume $c=\hbar=1$)
in order to solve the gravitational equations approximately
($\varphi$ is a typical gravitational
potential and $v$ a typical velocity in the system).
Observe, that $GT_{00}=G\rho=O(v^2)$ \cite{1} and,
therefore, the leading corrections for the
$g_{\mu\nu},\Psi_A$ are $O(v^2)$.
Thus, we can represent our fields in the form
\begin{equation}
g_{\mu\nu} = \eta_{\mu\nu}+h_{\mu\nu},\quad
\Psi_A = \Psi_A^0+\psi_A,\
\end{equation}
where $h_{\mu\nu},\psi_A=O(v^2)$.
We are interested in computing only the metric
since the test bodies are not sensitive to the other
fields.
It is well known \cite{1} that the first post-Newtonian
correction for the equations of motion of
massive test particles depends on  $h_{00}$ only.
This is just the Newtonian approximation and $-\frac12h_{00}$
is the gravitational potential.
But the post-Newtonian equations for light include both
the $h_{00}$ and $h_{ik}, i,k=1,2,3$ components
of the metric (the gravitational potential alone cannot
describe the motion of light!).

The $h_{\mu\nu},\psi_A$ can be obtained from
the linearized equations of the theory.
The invertible linear operator ${\cal D}$ of these equations
is constructed with the help of $\eta_{\mu\nu},\partial_\mu$
and $\Psi^0_A$. At this stage we work with the weak field
relativistic approximation. In the end we obtain the
required post-Newtonian approximation by dropping all
terms with time derivatives $\partial_0$ and replacing
$\Box$ by $\Delta$.
We assume that ${\cal D}$ does not contain
terms of the form $\Psi^0_\alpha\partial^\alpha$.
Such terms usually violate spatial isotropy, verified
experimentally with rather high accuracy \cite{1}.
Now we can use the spin projection operators \cite{11}
and decompose our equations on the independent spin sectors.
In general the field $h_{\mu\nu}$ can contain contributions
of 4 different kinds of particles
of spin $2^+,1^-,0^+,0^+$. The
corresponding projectors read
\begin{eqnarray}&&
P^{2^+}=\theta^\alpha_\mu\theta^\beta_\nu
  -\frac13\theta_{\mu\nu}\theta^{\alpha\beta},\
P^{1^-}=2\theta^\alpha_{(\mu}\omega^\beta_{\nu)},\
\nonumber\\&&
P^{0^+}_{1} = \frac13\theta_{\mu\nu}\theta^{\alpha\beta},\
P^{0^+}_{2} = \omega_{\mu\nu}\omega^{\alpha\beta},\
\nonumber\\&&
P^{0^+}_{12} = \theta_{\mu\nu}\omega^{\alpha\beta}/\sqrt{3},\
P^{0^+}_{21} = \omega_{\mu\nu}\theta^{\alpha\beta}/\sqrt{3},
\nonumber
\end{eqnarray}
here $\omega_{\mu\nu}=\partial_\mu\partial_\nu/\Box$,
and $\theta_{\mu\nu}=\eta_{\mu\nu}-\omega_{\mu\nu}$.
The subscripts 1,2 in the spin  $0^+$ sector label two different
kinds of particles of this type.
A similar spin decomposition exists for the field $\psi_A$
but we are not interested in its detailed contents
since the corresponding source $J_A$ vanishes.
Nonzero contribution to the metric for the source
$T_{00}=\rho$ can come only from $P^{2^+}T_{\mu\nu}$ and
$P^{0^+}_1T_{\mu\nu}$. All other projectors produce terms
which either vanish in the given approximation or
can be eliminated by an appropriate
choice of coordinate system. In general each
spin sector can contain several different particles.
The form of the linearized
equations in the spin $2^+$ sector is
\begin{equation}
\left(\begin{array}{ccc}
M_{00} P_0 & M_{01} P_{01} & \cdots \\
M_{10} P_{10} & M_{11} P_{1} & \cdots \\
\cdots & \cdots & \cdots
  \end{array}\right)
\left(\begin{array}{c} P_0 h_{\mu\nu} \\
  P_1\psi_A  \\ \vdots \end{array}\right)
=\left(\begin{array}{c} P_0 T_{\mu\nu}
  \\ 0 \\ \vdots \end{array}\right),\label{sector}
\end{equation}
where the coefficients $M_{ik}$ are {\em  scalar
polynomial functions} of the operator
$\Box=\partial_\alpha\partial^\alpha$ (here for simplicity we
omit the superscript $2^+$ and denote $P^{2^+}\equiv P_0$,
the subscripts 1,2,\dots label different $2^+$ modes).
The determinant of the matrix $M_{ij}$ has
the form $\prod (\Box-m^2_p)^q$.  The constants $m_p$ play the role of
``masses'' for the propagating linearized modes of our alternative gravity
theory.
Due to the orthogonality and completeness properties of
the spin projectors the solution of the equations
(\ref{sector}) can be obtained merely by calculating
the inverse matrix $N_{ik}=M^{-1}_{ik}$. We are interested to
know $N_{00}$ only; its general form
is $N_{00}=\sum\sigma_p/(\Box-m^2_p)$.
Finally we have to replace all
operators $1/(\Box-m^2)$ by $1/(\Delta-m^2)$
which leads to Yukawa exponential potentials.
The analysis of the spin $0^+$ sector is completely
analogous. Hence, we obtain a general form for
the metric
\widetext
\begin{eqnarray}&& 
g_{00} = -1+2\left[
(\sigma_0+\tau_0)U
+\sum^{n_2}_{p=1}\sigma_p U_p
+\sum^{n_2+n_0}_{q=n_2+1}\tau_q U_q
\right],\ \
\nonumber\\&& 
g_{ik} = \delta_{ik}\left(1+2\left[
(\frac{\sigma_0}2-\tau_0)U
+\sum^{n_2}_{p=1}\frac{\sigma_p}2 U_p
-\sum^{n_2+n_0}_{q=n_2+1}\tau_q U_q
\right]\right).\label{Gmetric}
\end{eqnarray} 
\narrowtext
\noindent Here the constants $\sigma_p, \tau_p$ depend on the
parameters of the concrete model and/or the constant
field $\Psi^0_A$. The constants $\sigma_p$ and $\tau_p$
represent the contributions of the spin $2^+$ and spin $0^+$
modes respectively. The Newtonian potential $U$
and exponential potentials $U_p$
\begin{equation}
U = G\int\frac{\rho'}
{{|\vec{x}-\vec{x}'|}}{d^3x'},\
U_p = G\int
\frac{\rho'{\rm e}^{-m_p{|\vec{x}-\vec{x}'|}}}
{{|\vec{x}-\vec{x}'|}}{d^3x'},\label{potentials}
\end{equation}
correspond to the massless and massive modes with
mass $m_p$.
Now all information about
a particular gravity theory is packed into several
constants $\tau_i,\sigma_k,m_p$.

We can compare the metric (\ref{Gmetric})
with the standard metric of the parametrized
post-Newtonian (PPN) formalism in the same
$O(v^2)$ approximation \cite{1}:
\begin{equation}
g_{00} = -1+2U,\qquad
g_{ik} =\delta_{ik}\left(1+2\gamma U\right).
\label{PPNmetric}
\end{equation}
Here the coefficient of $U$ in $g_{00}$
is fixed by the Newtonian limit and the experimental
value for $\gamma=1\pm10^{-3}$ comes from light
deflection and
retardation experiments in the solar system \cite{1}.
The only essential difference
between (\ref{Gmetric}) and (\ref{PPNmetric}) is that
the standard PPN formalism does not take into account a possible
contribution from massive modes.
Of course, the influence of massive exponential
potentials on the predictions of the gravity
model depends on the concrete values of the masses $m_p$.
If the mass of the mode is large enough, i.e., such that
$1/m_p$ is significantly less than 1cm, then the contribution
of $U_p$ can not be observed in gravity experiments
and our metric reduces effectively to (\ref{PPNmetric}).
On the other hand if $1/m_p$ is larger
than the typical size of a galaxy then
$U_p\approx U$ for galactic and shorter distances
and we are left once again with the metric (\ref{PPNmetric}).
In principle $1/m_p$ could be about the size of the Earth
or the solar system but in this case
experimental data impose very strong restrictions on the
magnitude of the constants $\sigma_p,\tau_p$ \cite{12}.
For example if $1/m_p \approx10^{13}{\rm cm}$
then $\sigma_p,\tau_p<10^{-8}$.

Therefore, we
have hopes to explain the dynamics of galaxies
if $1/m_p$ has an intermediate value significantly larger
than the size of the solar system but not greater than the typical size
of a galaxy (compare with (\ref{sanders})).
For our purposes it is sufficient to consider the simple case with
two massive particles
of spins $2^+$ and $0^+$ with approximately equal masses
$m_1\approx m_2\approx 10^{-26}{\rm eV}$.
Thus, we have (\ref{Gmetric}) with four unknown constants
$\sigma_0,\tau_0,\sigma_1,\tau_2$.
The experimental data imposes constraints on these parameters.
In the solar system $U_1\approx U_2\approx U$ and we have
\begin{equation}
\sigma_0+\tau_0+\sigma_1+\tau_2=1,\ \
{\textstyle\frac12}(\sigma_0+\sigma_1)-(\tau_0+\tau_2)=1,
\label{conditions-0}
\end{equation}
where the first condition ensures the correct Newtonian limit while
the second follows from experiments with light.
Now let us consider distances larger than 30kpc.
For these distances $U_1\approx U_2\approx0$.
As was mentioned above in order to reproduce the flat rotational
curves of galaxies without dark matter as far as is known
we only need the gravitational interaction to be
approximately 10 times stronger for large distances
\begin{equation}
\sigma_0+\tau_0\approx10,\quad
{\textstyle\frac12}\sigma_0-\tau_0\approx10.
\label{conditions-infty}
\end{equation}
Here the second conditions follow from
gravitational lensing, since
it is known \cite{2} that the observed
lensing is in conformity with the predictions
of Einstein's gravity {\em with dark matter}.
Thus, we have once again an approximately ten
times stronger
effective coupling constant for light.
Solving (\ref{conditions-0}),(\ref{conditions-infty})
one has
\begin{equation}
\sigma_0\approx\frac{40}3,\quad
\tau_0\approx-\frac{10}3,\quad
\sigma_1\approx-12,\quad
\tau_2\approx3\label{conditions}
\end{equation}
Any gravity model with the parameters (\ref{conditions})
should be in good correspondence with experimental data
in the solar system and should explain the behaviour
of stars and light in galaxies with reasonable accuracy.
Probably if we include more spin $0^+$ and $2^+$ particles
of various masses we could fit more details of the
galactic rotation curves \cite{8}.
This hypothetical model is not so simple:
the gravity theory should include, besides the
usual graviton, at least two extra massive very light
particles \cite{13}.
The most serious trouble with such a theory comes from
the negative sign of $\tau_0$ and $\sigma_1$.
In accordance with a general theorem \cite{14} the sign
of these constants must be {\em positive} for {\em even} spin
fields and {\em negative} for {\em odd} spin fields.
A wrong sign results in a propagating mode
carrying negative energy;
this is considered unacceptable in a theory.
There is still a small possibility of escaping
this problem. It has been suggested that if the theory
contains {\em several} particles with the same spin then
the conditions on the parameters
might be weakened \cite{8}.
In our opinion the possibility of successfully matching
the galactic rotation curves while avoiding negative energy
modes seems remote.

On the other hand, a negative sign for the coupling
constant is natural
for odd spin particles; initially it was
suggested that the exponential term in
(\ref{sanders}) is mediated
by a spin 1 vector particle \cite{9}.
In our scheme one can reproduce such a contribution
if and only if $J_A\neq0$. In particular, it was
suggested that $J_A$ may be proportional to baryonic
charge \cite{15}. However, the baryonic charge
to mass ratio varies from one body to another.
Therefore, such an interaction is no longer universal
and violates the weak equivalence principle \cite{16}
which has been verified experimentally
to a very high accuracy \cite{1}.
Now the importance and fundamental nature of postulate
(iii) becomes clear.  It ensures the universality
of the gravitational interaction.  All models which
satisfy this condition should not have trouble with
the equivalence principle at least
on the modern experimental level.

We have compared our model only with part
of the available experimental data.
Even if one overlooks the problem with the wrong sign of
the coupling constants the model must describe correctly,
in addition to the already considered effects,
the perihelion shift of Mercury,
the energy loss of the binary-pulsar,
and cosmological observations.
Comparison with these data cannot be performed
with the help of linearized equations and requires
a more detailed consideration for each particular
gravity theory. Of course, such comparison probably
result in {\em additional\/} perhaps severe
restrictions on the theory under consideration
(see e.g. \cite{add} which discuss cosmological
restrictions).

Although our scheme covers a large class of the theories
yet we can not exclude the possibility of constructing
a model with the desired properties
which violates one of our postulates.
We want to mention briefly some possibilities which
have been proposed.

1. It has been shown \cite{17} that a cosmological constant
$\Lambda \approx 10^{-52}{\rm cm}^{-2}$ is able to
explain the flat rotational curves of galaxies.
This value of $\Lambda$ is ten times bigger than
the cosmologically acceptable limit.

2. Consideration of quantum corrections to the
Newtonian potential may result in some additional
logarithmic long range terms \cite{18}.
The applicability of these results to galactic
distances is not obvious.

3. A quadratic in curvature lagrangian
can also produce modified gravitational
potentials with extra long range terms \cite{7}.
This model has higher derivatives equations and
requires a traceless energy-momentum tensor.

4. It is possible to construct an essentially nonlinear
theory with a non-quadratic kinetic term in the Lagrangian.
This model explains the dynamics of galaxies due to
the nonlinear nature of the equations in the regime
of small accelerations \cite{16,5}.  It was
observed that such a theory may have troubles with
faster-than-light waves.

Although all these models explain the rotation of
galaxies they have their own weak points, moreover they
must be capable of predicting correctly other
gravitational effects.
Therefore we conclude that the possibility of explaining
galactic mysteries with the help of a modified gravity
theory looks quite improbable.

This work has been supported by the
National Science Council of
the Republic of China under contracts No
NSC 83-0208-M008-014 and NSC 83-0208-M008-028.
We thank the referee for informing us of Ref. \cite{add}.


\end{document}